\documentclass[useAMS,usenatbib,usedcolumn,usegraphicx]{mn2e}
\def\etal{{\it et\thinspace al.}\ }
\def\pii{{{\sc P\,ii}}\ }
\def\piii{{{\sc P\,iii}}\ }

\def\mum{{$\mu$m}\ }

\def\eion{{(e~+~ion)}\ }

\usepackage[dvips]{graphics}
\usepackage[dvipsnames,usenames]{color}
\usepackage{lscape}

\newcommand{\be}{\begin{equation}}
\newcommand{\ee}{\end{equation}}
\title{Biosignature Line Ratios of [P II] 
in Exoplanetary and Nebular Environments}
\author
[Kevin Hoy, Sultana N. Nahar, Anil K.\ Pradhan]
       {Kevin Hoy$^1$, Sultana N. Nahar$^1$, Anil K.
Pradhan$^{1,2,3}$\\
       $^1$ Department of Astronomy, $^2$ Chemical Physics Program, $^3$
Biophysics Graduate Program,
 The Ohio State University, Columbus, OH 43210, USA.}
\date{Accepted  xxxxxx 
      Received xxxxxx;
      in original form xxxxxx}

\pagerange{\pageref{firstpage}--\pageref{lastpage}}
\pubyear{2022}

\def\LaTeX{L\kern-.36em\raise.3ex\hbox{a}\kern-.15em
    T\kern-.1667em\lower.7ex\hbox{E}\kern-.125emX}

\begin{document}

\maketitle

\label{firstpage}

\begin{abstract}
    Being the backbone element of DNA, phosphorus is a key component in the 
search for life in the Universe. To aid in its detection, we present 
line emissivity ratios for the five lowest-lying forbidden [P~II] transitions, 
namely those among the levels $3s^23p^2(^3P_0,^3P_1, ^3P_2,^1D_2,^1S_0)$. 
The wavelengths range between 0.44-70 \mum, and several lie within the
spectroscopic domain observable with the James Webb Space Telescope
(JWST). 
These line ratios have been calculated using a new
collisional-radiative-recombination (CRR) model combining calculated collision 
strengths and level-specific recombination rate coefficients; with both datasets
computed using the accurate Breit-Pauli R-Matrix method. 
The CRR model includes a new scheme for \eion recombination to emission
line formation. We compare its effect to
models incorporating only electron impact excitation and spontaneous 
radiative decay. 
We find that \eion recombination has a significant impact on all line ratios, 
and represents a major improvement in physical accuracy of
emission line models.

\end{abstract}


%\section{Equations}
% Line Ratio:
%\be \frac{I_{jk}}{I_{mn}} = \frac{P_{jk}}{P_{m}} \frac{A_{jk}}{A_{mn}} 
%\frac{E_{jk}}{E_{mn}}  \ee

%Cascade Coefficient:
%\be Cji = \frac{A_{ji}}{\sum_{i'<j} A_{ji'}} + \sum_{k>i} D_{jk} \frac{A_{ki}}{\sum_{i'<k}A_{ki'}} \ee

%RRC:

%$$\alpha_R(T) = \frac{g_j}{g_s} \frac{2}{kT\sqrt{2\pi m^3c^2kT}} \int_0^\infty (I+\epsilon)^2 \sigma_{PI}(\epsilon)e^\frac{-\epsilon}{kT}d\epsilon $$
%\be = 1.8526 \times 10^4 \frac{g_j}{g_s} \int_0^\infty (\epsilon + I)^2 e^\frac{-\epsilon}{kT}\sigma_{PI}d\epsilon \: cm^3 s^{-1}  \ee

\section{Introduction}
Phosphorus could be of key importance in the search for life in the Universe. 
As the backbone element of DNA, it could be a fundamental prerequisite for the formation of life. 
It is thus of keen interest to perform theoretical calculations that can provide a basis of comparison for real observations of exoplanets.
In order to determine if phosphorus is present in an observed exoplanet, we must first understand the elements spectral signature.
To that end, it is necessary to build an accurate model of as many physical processes that influence the emission spectrum of phosphorus as possible. 

 Previous studies on phosphorus abundance have been confined to ionized
sources such as nebulae (Pottsch \etal 2008,
Pottasch and Bernard-Salas 2008, Otsuka \etal 2011), and
FGK dwarf stars (Maas \etal 2017). The former are associated with
star-forming regions, and the latter are potential sources for
Earth-like extra-solar planets. Nucleosynthesis leading to phosphorus
(Z=15) production can occur in AGB stars, but mainly in supernovae by neutron
capture with silicon (Z=14). Koo \etal (2013) analyzed the P/Fe abundance
ratio in the supernova remnant Cassiopia A using forbidden \pii 1.189
\mum line and {\sc Fe\,ii} 1.257 \mum lines to deduce about 100 times the
average ratio in the Milky Way. Otherwise, phosphorus abundances are generally
low in solar system  $2.6 \times 10^{-7}$ by number relative to
hydrogen (Asplund \etal 2009). 
Also, its gas phase abundance is difficult to determine due to
dust and grain formation since phosphorus is very reactive.

Of all biosignature elements H, C, N, O, P and S, atomic data for
P-ions is the least available, except a limited number of 
energies from the National
Institute for Standards and Technology (NIST), 
theoretial photoionization and \eion recombination data (Nahar 2016,
2017), experimental measurements and comparison with theory for \pii
(Hernandez \etal 2015, Nahar \etal 2017), and collision strengths for
\piii (Naghma \etal 2018).
With the aim of more comprehensive analysis of observational data, 
we compute the relevant atomic parameters and line intensity ratios for
\pii. In particular, we study lines that might be detectable with JWST
in the NIR range 0.6-28 \mum.
The primary focus of this work is
the description of a major improvement to modelling emission lines by
the inclusion of detailed and accurate
\eion recombination contribution to forbidden lines
from recombination-cascades via high-lying levels up to $n \leq 10$, 
and its application to the modelling of \pii emission.
Generally, we consider collisional-radiative models wherein 
electron impact excitation is the primary means of 
populating low-lying excited levels which produce forbidden emission
lines.
In this work, we generalize the model to a collisional-radiative-cascade
(CRR) model in which \eion recombination followed by myriad radiative
pathways significantly populate upper levels.
This additional contribution yields considerable change in level
populations and line emissivities, demonstrating the importance of 
utilizing models with \eion recombination and radiative cascades.
The treatment described is generally applicable to modeling atomic
spectra.

\section{Theory \& Computational Method}
A theoretical description of key atomic parameters is outlined here.

\subsection{Energy Levels \& Einstein A Coefficients}
In order to calculate the necessary energy levels and 
Einstein A-coefficients for spontaneous radiative decays, we have employed 
the atomic structure code SUPERSTRUCTURE (Eissner \etal 1974). 
The code operates using the Breit-Pauli approximation, and produces fine
structure level energies and associated radiative transition rates.
Figure~1 presents the Grotrian energy level diagram for the 5 
lowest fine structure levels of \pii. 
All levels presented are within the ground configuration $1s^22s^23s^23p^2$. 
The energy difference between the ground state $^3P_0$ and the highest
level $^1S_0$ is about 0.2 Ryd or 2.72 eV.
All levels are of even parity, thus all transitions between them and
discussed in this work are forbidden 
due to electric quadrupole E2 
and magnetic dipole M1 second order relativistic interactions.

\begin{figure} %figure using just one column
\centering  
\includegraphics[width=\columnwidth,keepaspectratio]{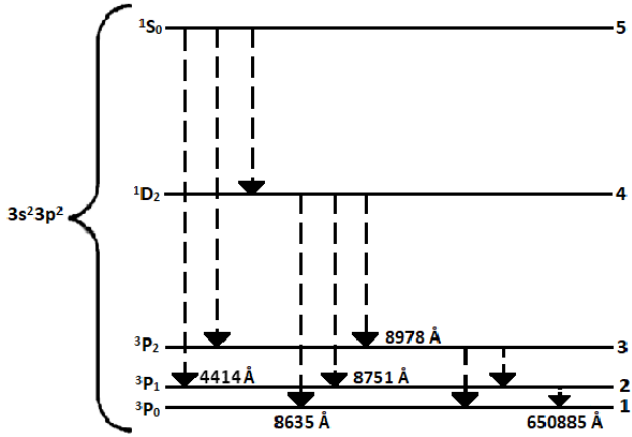}
\caption{Energy diagram of {\sc P\,ii} showing the fine structure levels
in the ground configuration $3s^23p^2$, and a few transitions in the JWST
wavelength range. 
\label{energies}} 
\end{figure}

\subsection{Electron Impact Excitation}
When considering the effects of of Electron Impact Excitation there are 
two quantities to consider. First, the collision strengths which are 
related to the probability and cross section as function of energy
of an incident electron at a given energy E causing an excitation 
from a given initial level $i$ to a final level $j$. 
For astrophysical plasma sources we generally assume a Maxwellian
distribution of electron energies correspnding to a given electron
kinetic temperature $T_e$. Therefore,
for calculating line emissivity ratios, we compute and employ Maxwellian 
averaged effective collision strength $\Upsilon(T_e)$
at a range of $T_e$ characteristic
of the source, and that is related to the rate at which a given transition is
collisionally excited,

\be \Upsilon_{ij}(T_e) = \int_0^\infty \Omega_{ij} (E) \exp(-E/kT_e) d(E/kT_e). \ee

\noindent Here, $\Omega_{ij}(E)$ is the collision strength 
for the transition $i \rightarrow j$; $k$ is the Boltzman constant. 
Fig.~2 shows a sample of computed collision strengths and 
effective collision strengths. As seen, 
collision strengths demonstrate extensive 
resonance structures which make them highly dependent on electron energy, 
while effective collision strengths are slowly varying with $T_e$,
also implying that slight errors in temperature have a less significant 
impact on any derived result.

Effective collision strengths have been calculated for 
the 67 lowest lying \pii levels. We also compare the computed line
ratios from the present CRR model using previously calculated \pii
collision strengths (Tayal 2004). We have employed level-specific 
recombination rate coefficients (Nahar 2017a, 2017b) for 243 levels.
For levels not included in the models
the effective collision strengths are set to zero.
The inclusion of many more levels for \eion recombination-cascades 
should offset the insignficant loss of accuracy from omitting the 
electron impact excitation to higher levels.

\subsection{Electron-ion recombination}

The contribution of recombination is contained in two key parameters,
recombination rate coefficients and cascade coefficients. Level-specific
recombination rates refer to electron recombining
with an ion, in this case \piii to form \pii in an excited
state. They are defined by Eq.~2, where g$_j$ is the statistical
weight factor of the final state of the ion being recombined, g$_s$ is
the statistical weight of the initial state of the recombining ion, k is
the Boltzmann constant, T is the temperature of the plasma, m is the
mass of the electron, c is the speed of light, $\epsilon$ is the energy
of the incident electron, and $\sigma_{PI}(\epsilon,j$) 
is the photoionization cross section a given electron energy. 
The Milne relation gives the level-specific rate coefficient into level $j$
as:

%\be
\begin{eqnarray}
\alpha_R(j,T) & = & \frac{g_j}{g_s} \frac{2}{kT\sqrt{2\pi m^3c^2kT}}
%\alpha_R(j,T)  =  \frac{g_j}{g_s} \frac{2}{kT\sqrt{2\pi m^3c^2kT}}
\int_0^\infty (I+\epsilon)^2 \sigma_{PI}(\epsilon,j)
e^\frac{-\epsilon}{kT}d\epsilon\\
& = & 1.8526 \times 10^4 \frac{g_j}{g_s} \int_0^\infty (\epsilon + I)^2
% =  1.8526 \times 10^4 \frac{g_j}{g_s} \int_0^\infty (\epsilon + I)^2
e^\frac{-\epsilon}{kT}\sigma_{PI}(\epsilon,j)d\epsilon \: cm^3
s^{-1}. \nonumber
%s^{-1}.
\end{eqnarray}
%\ee

\noindent The cascade coefficients C$_{ji}$ are calculated from BPRM
A-values and used to obtain the fractional contribution $j \rightarrow i$
due to \eion recombination into an upper level $j$
to a lower level $i$. The C$_{ji}$  take into account all spontaneous 
radiative decays via dipole E1 transtions through 
intermediate levels $k$ such that $i < k < j$ as follows, 

\be C_{ji} = \frac{A_{ji}}{\sum_{i'<j} A_{ji'}} + \sum_{k>i} C_{jk} \frac{A_{ki}}{\sum_{i'<k}A_{ki'}} \ee

\noindent The A-values are the only necessary quantities for 
calculating cascade coefficients; C$_jk$ are subsets of 
intermediate cascade coefficients, and combining these
we obtain the total effective recombination contribution to a given level
population.

\subsection{Line Emissivities \& Ratios}
Combining the results above we obtain line emissivity ratios from CRR models.
The theoretical line emissivity is the amount of energy per unit time per unit
volume for a given transition. For a given \pii transition $j \rightarrow i$ 
it may be expressed as,

$$I([PII]; \lambda_{ji}) = \frac{h\nu_{ji} A_{ji}}{4\pi}\times
\frac{N_j}{\sum_k N_k} \times \frac{n(PII)}{n(P)}$$ 
\be \times \frac{n(P)}{n(H)} \times n(H) \; erg \; cm^{-3} \; s^{-1}. \ee

\noindent $N_j$ refers to the population of the upper state. Summation 
over $N_k$ refers to the total population of all levels included in the
calculation, and n(\pii)/n(P) is the density ratio relative to
phosphorus abundance P/H.
In order to obtain line emissivities form Eq.~4 one needs ionization
fractions \pii/P in the plasma source at the appropriate temperature and
density, which are generally unknown {\it a priori} and are model
dependent. 
We compute line ratios in a temperature-density range of exoplanetary
ionospheres as

\be \frac{I(PII; \lambda_{ji})}{I(PII; \lambda_{mn})} = \frac{N_j A_{ji}
\nu_{jk}}{N_m A_{mn} \nu_{mn}}, \ee

for any two transitions $j \rightarrow i$ and $m \rightarrow n$.
The ratio in Eq.~5 eliminates the number density factors and instead 
leaves only the relative population factor, Einstein A values, 
and energies of the two transitions being considered. 
The key quantities here are relative populations of upper levels
computed by our CRR models.

\subsection{Code SPECTRA}

These results have been produced by a new version of the 
code SPECTRA, employed in many past calculations of line ratios. Previously, the code considered electron impact excitation and 
photoexcitation for populating excited levels. The updated version
accounts for \eion recombination as described. Moreover, the new code is
written in C++ rather than Fortran which increases its utitliy,
efficiency, and portability. 

The computational work
was performed in two distinct stages.
First, there was a direct unaltered translation from Fortran to C++.
During this stage, the primary goal was to produce matching results with
identical input data to many significant figures.
The second stage included the recombination-cascade formalism described
herein.
As there are no previous results to compare with, extensive testing was
performed and checked several different ways (including manually) to
ensure accuracy and reliability for modelling emission lines with
recombination contributions in general.

\section{Results and discussion}
The main results presented are the newly calculated collision strengths, 
Maxwellian-averaged effective collision strengths, and line emissivity ratios.
The former two have been computed for all transitions among the first 67
\pii levels, with a small sample of select results presented in table 1,
and shown in Fig.~2 for two transitions $^3P_1-^1D_2$ and $^3P_1-^1S_0$
at wavelengths 8751$\AA$ and 4413$\AA$, respectively.
The emissivity ratios are presented in Figures 3 \& 4. 

\subsection{Effective Collision Strengths}
The collision strengths and effective collision strengths for two select transitions can be seen in Figure 2. 
The strong resonant structures in the former are clearly visible, as is the signature smoothing of the effective collision strengths. 

%\begin{figure*}
%\begin{minipage}{148mm}
\begin{figure} %figure using just one column
\centering  
\includegraphics[width=\columnwidth,keepaspectratio]{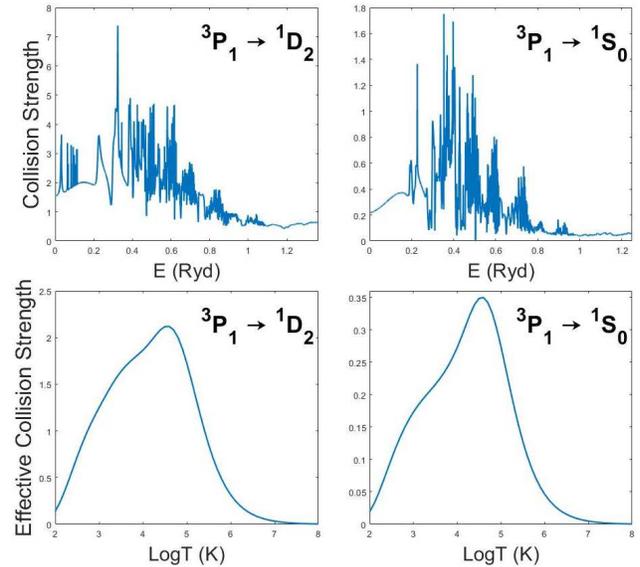}
\caption{Collision Strengths vs. E(Ryd) (upper panels), and Maxwellian
averaged effective collision strengths vs. Log T(K) (lower panels)
for $^3P_1-^1D_2)$ (left) and $^3P_1-^1S_0)$ (right).
The Collision strengths are dominated by autoionizing resonances in the
near-threshold energy region, whereas the 
effective collision strengths vary smoothly with temperature but
strongly peaked around $10^4$K.
\label{fir}} 
%\end{minipage}
\end{figure}

 The temperature dependence of effective collision strengths due to
autoionizing resonances is quite pronounced, and is different from that
of \eion recombination. The competition between the two
processes populating upper levels and leads to very different line emissivity 
ratios. 

\subsection{Line Emissivity Ratios}
Line ratios for select transitions among the first 5 \pii levels can be seen in Figure 3. 
The key features to note are the differences among the 
three independent curves in each plot with inclusion of \eion
recombination and without. 
We find a large increase with \eion recombination-cascades in the CRR
model (red curves), and much lower
emissivity ratios without (blue and black curves) using
our newly calculated collision strengths and earlier work by Tayal
(2004), respectively. 
Although the two sets of collision strengths data yield practically the
same values
at low densities $N_e < 10^5$ cm$^{-3}$, the differences increase at
higher densities.
That is owng to collisional coupling among the 5
levels which becomes more prominent and manifested in line ratios.  
Tayal's collision strengths are higher than present
ones for excitation of upper levels considered and yield 
higher line ratios at $N_e > 10^5$ cm$^{-3}$; those 
would be even higher if \eion recombination is included.
This illustrates not only a significant improvement in 
accuracy using the CRR model, but also that \eion recombination using
detailed and precise level-specific recombination rates should be used
as basis of comparison with observed emission-line spectra. 

\begin{figure} %figure using just one column
\centering  
\includegraphics[width=\columnwidth,keepaspectratio]{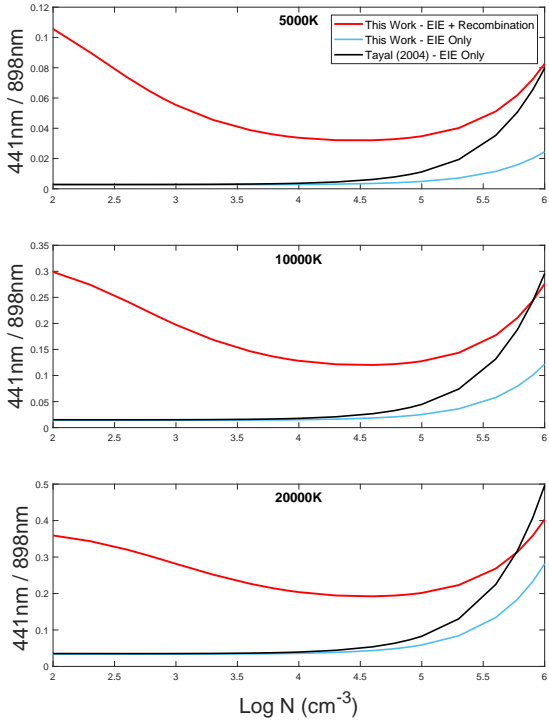}
\caption{[\pii] line emissivity ratios as function of temperature and density
demonstrating the effects 
of newly calculated collision strengths and the incorporation of \eion
recombination-cascades. The red curve represents 
results including recombination and present collision strengths. 
The black and blue curves are calculated without recombination using
present collision strengths (blue) and those from Tayal (2004; black), 
respectively. 
%In each of the three plots, the numerator is held constant to the lowest transition, $^3P_1-^3P_0$, with the denominator changing to the labelled transition. 
\label{energies}} 
\end{figure}

Figure 4 includes line ratios for select transitions whose wavelengths are 
within JWST spectrographic instrumental range 0.6-28\mum at the three
different temperatures. However, in
this case the upper level $1D_2$ is the same for all ratios and
therefore line emissivities depend only on the respective A-values, and
to a smaller extent on the energy differences. The A-values for the
three transitions $1D_2 \rightarrow 3P_{0,1,2}$ are: $2.177 \times
10^{-5}, \ 6.744 \times 10^{-3}, \ 1.938 \times 10^{-2}$ sec$^{-1}$,
respectively. Therefore, the {\it proportion} of line ratios 
0.898$\mu$m/0.890$\mu$m and
0.875$\mu$m/0.890$\mu$m due to $^1D_2-^3P_2$/$^1S_0$ and 
$^1D2-^3P_1$/$^1S_0-^1D2$ remains
approximately constant 3:1 (modulo the small energy differences).
Again, we find large differences upon inclusion of recombination-cascade
contributions to emission lines owing to upper levels being populated
predominantly by \eion recombination rather than excitation due to 
electron impact. 

 As the electron density increases above $N_e >
10^4$ cm$-3$, recombination-cascades are relatively less important
compared to electron collisions and the two curves are
eventually seen to converge at high densities, though the proporation
remains the same as it should owing to origin of line emissivity from
a single upper level.
We note in passing that there are several transitions orginating from
the same level $^1D_2$ where the line ratios show similar behavior and
are constant across temperatures and densities since
only the upper level's population is sensitive to either and
present in both numerator and denominator. As such line ratios depend
only on ratios of A-values and energy differences, and provide a  
an accuracy check on input data.

\begin{figure} %figure using just one column
\centering  
\includegraphics[width=\columnwidth,keepaspectratio]{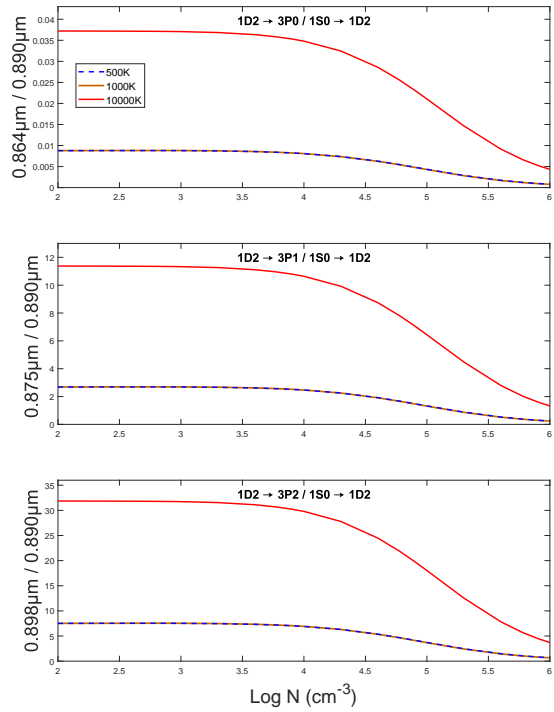}
\caption{Emissivity ratios of [\pii] lines at ionospheric and nebular
temperatures 500K, 1000K, and 10,000K lying within and potentially
observaby by JWST spectrometers. 
\label{energies}} 
\end{figure}

Line emissivities for [\pii] transitions within the ground configuration
are computed using two sets of collision strengths of data, computed as
part of the present work as as previously reported in literature (Tayal
2004). Although the line ratios are in good agreement,
they differ significantly; the new values should be more accurate. 
The CRR models can be further improved by the consideration of 
fluorescent excitation due to an external source of radiation, 
namely the planet's host star.
SPECTRA maintains the ability to account for that process and can 
incorporate that mechanism, though it is likely to be much smaller than
electron excitation and \eion recombination-cascade processes for
forbidden line formation.

\begin{table*}
%\begin{minipage}{148mm}
\begin{minipage}{188mm}
\caption{Effective Maxwellian averaged collision strengths for select transitons in {\sc P\,ii}}
\begin{tabular}{cccccccccc}
%\begin{tabular}{llllllllll}
\hline
%$LogT$(K) & $^2P^o_{1/2-3/2}$ & $^2P^o_{1/2}-^2D_{3/2}$ & $^2P^o_{3/2}-^2D_{3/2}$ &$^2P^o_{3/2}-^2D_{5/2}$ & 
%$LogT$(K) & $^2P^o_{1/2-3/2}$ & $^2P^o_{1/2}-^2D_{3/2}$ & $^2P^o_{3/2}-^2D_{3/2}$ &$^2P^o_{3/2}-^2D_{5/2}$\\
$N_e(cm^{-3}$) & $^3P-^1D$ & $^3P-^1D$ & $^3P-^1D$ &$^1D-^1S$ & 
$N_e(cm^{-3}$) & $^3P-^1D$ & $^3P-^1D$ & $^3P-^1D$ &$^1D-^1S$\\
 J-J' & 0-2 & 1-2 & 2-2 & 2-0 & J-J' & 0-2 &
1-2 & 2-2 & 2-0\\
$\lambda$ &  8862.6 $ \AA$ &   4441.7 $\AA$ & 2916.7 $\AA$ & 1571.2 $\AA$
& $\lambda$ &  8862.6 $ \AA$ &   4441.7 $\AA$ & 2916.7 $\AA$ & 1571.2 $\AA$ \\
\hline

	2.0& 5.04(-1)&	 1.40(-1)&  2.97(-3)& 	2.01(-56)&   
    3.6& 6.11(-1)&	 1.693&	    2.552&	    2.91(-2)\\
	2.1& 5.07(-1)&	 2.23(-1)&  1.20(-2)&	5.29(-45)&
    3.7& 6.19(-1)&	 1.740&	    2.683&	    5.77(-2)\\
	2.2& 5.10(-1)&	 3.39(-1)&  3.62(-2)&	6.23(-36)&
    3.8& 6.27(-1)&	 1.781&	    2.798&	    1.01(-1)\\
	2.3& 5.12(-1)&	 4.63(-1)&  8.73(-2)&	1.00(-28)&
    3.9& 6.34(-1)&	 1.822&	    2.902&	    1.59(-1)\\
	2.4& 5.14(-1)&   5.94(-1)&  1.76(-1)&	5.29(-23)&
    4.0& 6.45(-1)&	 1.866&	    3.005&	    2.32(-1)\\        
	2.5& 5.16(-1)&	 7.23(-1)&  3.06(-1)&	1.86(-18)&
    4.1& 6.57(-1)&	 1.915&	    3.113&	    3.21(-1)\\
	2.6& 5.17(-1)&	 8.47(-1)&  4.77(-1)&	7.62(-15)&
    4.2& 6.72(-1)&	 1.972&	    3.228&	    4.25(-1)\\
	2.7& 5.20(-1)&	 9.61(-1)&  6.78(-1)&	5.64(-12)&
	4.3& 6.89(-1)&	 2.031&	    3.343&	    5.41(-1)\\
	2.8& 5.23(-1)&	 1.065&     8.99(-1)&	1.07(-9)&
	4.4& 7.03(-1)&	 2.084&	    3.444&	    6.65(-1)\\
	2.9& 5.28(-1)&	 1.160&     1.129&	    6.94(-8)&
    4.5& 7.11(-1)&	 2.117&	    3.511&	    7.89(-1)\\
	3.0& 5.36(-1)&	 1.250&     1.362&	    1.91(-6)&
	4.6& 7.10(-1)&	 2.119&	    3.524&	    9.00(-1) \\
	3.1& 5.47(-1)&	 1.337&     1.593&	    3.00(-5)&
	4.7& 6.95(-1)&	 2.081&	    3.469&	    9.91(-1) \\
	3.2& 5.60(-1)&	 1.421&     1.817&	    2.20(-4)&
  	4.8& 6.66(-1)&	 2.000&	    3.340&	    1.050 \\
	3.3& 5.75(-1)&	 1.501&     2.030&	    1.14(-3)&
	4.9& 6.25(-1)&	 1.879&	    3.143&	    1.071 \\
	3.4& 5.89(-1)&	 1.574&     2.226&	    4.31(-3)&
    5.0& 5.73(-1)&	 1.725&	    2.890&      1.054 \\
	3.5& 6.01(-1)&	 1.638&	    2.400&	    1.24(-2)&&&&\\
	
\hline
\end{tabular}
\end{minipage}
\end{table*}

\section{Conclusion}
The computed data and results are useful for both temperature/density 
diagnostics and possible detection of phosphorus in astrophysical sources,
particularly exoplanetary ionosopheres as biosignatures or precursors of
DNA-based life forms.
Line emissivity ratios and new collision strengths for \pii are computed for 
comparison with observation and diagnostics. 
The inclusion of \eion recombination 
has a large impact on forbidden line emissivities, and considerably
improve the physical accuracy of models.
Many transitions, with a sample of results presented, 
occur within the 0.6-28.3 \mum range of the spectrometers aboard JWST. 
Given the sensitivity of JWST, these emission lines might 
be present in the spectra of exoplanets that have detectable amounts of 

Future work will involve a more complete CRR modelling of resultant
spectrum with SPECTRA involving emission and absorption processes. 
This entails user input of ionization fractions and abundances
to produce individual line emissivities, in addition to line ratios.
Data from this new code can then be combined with codes for a full-scale
modelling of planetary atmospheres, such as a new version
of the code GEANT4-EXOP code in progress (M. Rothman, private
communication). These models will include atomic-ionic-molecular data as
input to produce absorption and emission spectra of other biosignature
elements.

\section*{Acknowledgments}
 The computational work was 
carried out at the Ohio Supercomputer Center in Columbus 
Ohio. 

\section*{Data Availability}

 All data underlying the results presented are available publicly from
the online database Nahar-OSU-Atomic-Radiative-Data (NORAD), maintained
by co-author S.N. Nahar at the Department of Astronomy of the Ohio State
University from the website https://norad.astronomy.ohio-state.edu.

\label{lastpage}

\frenchspacing
\def\aa{{\it Astron. Astrophys.}\ }
\def\aasup{{\it Astron. Astrophys. Suppl. Ser.}\ }
\def\adndt{{\it Atom. data and Nucl. Data Tables.}\ }
\def\aj{{\it Astron. J.}\ }
\def\apj{{\it Astrophys. J.}\ }
\def\apjs{{\it Astrophys. J. Supp. Ser.}\ }
\def\apjl{{\it Astrophys. J. Lett.}\ }
\def\baas{{\it Bull. Amer. Astron. Soc.}\ }
\def\cpc{{\it Comput. Phys. Commun.}\ }
\def\jpb{{\it J. Phys. B}\ } 
\def\jqsrt{{\it J. Quant. Spectrosc. Radiat. Transfer}\ }
\def\mn{{\it Mon. Not. R. astr. Soc.}\ }
\def\pasp{{\it Pub. Astron. Soc. Pacific}\ }
\def\pra{{\it Phys. Rev. A}\ }
\def\pr{{\it Phys.  Rev.}\ } 
\def\prl{{\it Phys. Rev. Lett.}\ }

\bibliography{ms}
\end{document}